\documentclass[11pt]{article}
\usepackage{amssymb,amsmath,cite,geometry,graphicx,url}
\catcode`\@=11

\@addtoreset{equation}{section}
\def\theequation{\arabic{section}.\arabic{equation}}
     
\catcode`\@=11
\def\thesection{\arabic{section}.}

\def\appendix{\setcounter{section}{0}
        \def\thesection{Appendix.}
        \def\theequation{\Alph{section}.\arabic{equation}}}
\def\section{\@startsection{section}{1}{\z@}{3.5ex plus 1ex minus
   .2ex}{2.3ex plus .2ex}{\large\bf}}
     
\long\def\@makefntext#1{\parindent 0cm\noindent
\hbox to 1em{\hss$^{\@thefnmark}$}#1}

 \makeatother   

\begin{document}
\begin{titlepage}
\vspace{.5in}
\begin{flushright}
March 2019\\  
\end{flushright}
\vspace{.5in}
\begin{center}
{\Large\bf
 Dimension and Dimensional Reduction\\[1ex] in Quantum Gravity}\\  
\vspace{.4in}
{S.~C{\sc arlip}\footnote{\it email: carlip@physics.ucdavis.edu}\\
       {\small\it Department of Physics}\\
       {\small\it University of California}\\
       {\small\it Davis, CA 95616}\\{\small\it USA}}
\end{center}

\vspace{.5in}
\begin{center}
{\large\bf Abstract}
\end{center}
\begin{center}
\begin{minipage}{4.85in}
{\small
If gravity is asymptotically safe, operators will exhibit 
anomalous scaling at the ultraviolet fixed point in a way that makes the
theory effectively two-dimensional.  A number of independent lines 
of evidence, based on different approaches to quantization, indicate
a similar short-distance dimensional reduction. I will review the 
evidence for this behavior, emphasizing the physical question of 
what one means by ``dimension'' in a quantum spacetime, and will 
discuss possible mechanisms that could explain the universality of 
this phenomenon.\\[2ex]

{\small\it For proceedings of the conference in honor of Martin Reuter:\\[.5ex]
``Quantum Fields---From Fundamental Concepts to Phenomenological Questions''\\[.5ex]
September 2018}
 }
\end{minipage}

\end{center}
\end{titlepage}
\addtocounter{footnote}{-1}
 
\section{Introduction}
The asymptotic safety program offers a fascinating possibility for the quantization of
gravity, starkly different from other, more common approaches, such as string theory and
loop quantum gravity \cite{Weinberg,Reuter,AS}.  We don't know whether quantum 
gravity can be described by an asymptotically safe (and unitary) field theory, but it might 
be.  For ``traditionalists'' working on quantum gravity, this raises a fundamental question: 
What does asymptotic safety tell us about the small-scale structure of~spacetime?

So far, the most intriguing answer to this question involves the phenomenon of
short-distance dimensional reduction.  It seems nearly certain that, near a nontrivial
ultraviolet fixed point, operators acquire large anomalous dimensions, in such a way
that their effective dimensions are those of operators in a two-dimensional spacetime
\cite{Reuterb,Niedermeyer,Reuterc}.  As the clocks and rulers we use to measure
spacetime are made up of quantum fields, this means that the dimension of spacetime 
is effectively reduced to $d=2$.  

If this behavior were unique to the asymptotic safety program, it would tell us something
important about asymptotic safety.  However, the same kind of dimensional reduction shows up
in a wide variety of conceptually very different approaches to quantum gravity.  Indeed,
dimensional reduction was first noticed in high temperature string theory as early as 1988
\cite{Atick}, and has reappeared in discrete approaches (causal dynamical
triangulations \cite{Loll}, causal sets \cite{Carlip1}), loop quantum \mbox{gravity \cite{Modesto}},
\mbox{the Wheeler-DeWitt} equation \cite{Carlip2}, noncommutative geometry \cite{Nozari},
and a broad range of other research programs.  Perhaps this is telling us not only about
asymptotic safety, but about some even more fundamental feature of short-distance
spacetime.

In this article, I will review some of the hints of dimensional reduction, examine possible
physical mechanisms, and discuss the possibility of experimental tests.  First, though, we
will have to address a fundamental point: How does one define a ``dimension'' as an observable?

Much of the work I describe here is based on the review article \cite{Carlip3}, to which I
refer the reader to, for more details.
 
\section{What Do We Mean by ``Dimension''?}

How do we know  we live in a three-dimensional space?  Here's one way:  Pick a point $P$,
and draw a straight line through it.  Draw a second line through $P$ perpendicular to the first,
and a third perpendicular to the first two.  That's the best we can do; there is no fourth line
perpendicular  to the first three.  

This demonstration goes back at least to Galileo \cite{Gal}, and perhaps to a lost manuscript by
Ptolemy (see \cite{Ptol}).  However, it clearly contains assumptions, both mathematical and physical: We must
have well-defined notions of ``straight'' and ``perpendicular'', and our eyes and pencils must have
access to all dimensions.  The construction can be broadened to the more general statement that
three coordinates are needed to locate a point in space; but, with Cantor's discovery of a one-to-one
map between a line segment and a region of $\mathbb{R}^n$ and Peano's construction of space-filling 
curves \cite{Hurewicz},  \mbox{the simple} picture of dimension as a count of parameters became inadequate.  
Indeed, it wasn't until 1911 that it was even proven that $\mathbb{R}^m$ and $\mathbb{R}^n$,  $m\ne n$, were 
not homeomorphic \cite{Brouwer}.

Meanwhile, physicists were also struggling with parallel questions about the nature of 
dimension.  In a famous 1917 paper, Ehrenfest pointed out that features such
as the stability of orbits and electromagnetic duality required three spatial dimensions, but
added that ``the questions [of what determines the number of dimensions] have perhaps no 
sense'' \cite{Ehrenfest}.  The works of Nordstrom \cite{Nordstrom}, Kaluza \cite{Kaluza}, and Klein
\cite{Klein} introduced the idea of ``hidden'' dimensions, while, much later, string theory suggested
that perhaps spacetime dimension could be determined by some underlying principle.

The mathematicians' reaction to this crisis was the development of ``dimension theory'' as a
branch of topology, with several rigorous and general definitions of dimension: inductive
dimension (\mbox{a point} divides a line into two pieces, a line divides a planar region, a plane divides
a three-dimensional region, etc.) and covering dimension (to cover a line, pairs of open sets must
overlap; to cover a planar region, triples of open sets must overlap; and so on) \cite{Hurewicz}. However, these 
are perhaps too primitive for physics, relying on concepts such as ``open sets'' that are not so easy 
to determine in the physical~world.  

Physicists have, therefore, come to depend on a collection of other definitions of dimension, some
of them more mathematical and others more directly tied to physical quantities.  These may be
best thought of as ``dimensional estimators'', observables whose values correlate with something
we might think of as the dimension of spacetime.  Crucially, these estimators need not always agree:
\mbox{The dimension} of spacetime may depend on how one chooses to ask the question.  Some of the
most common choices are the following (see \cite{Carlip3} for details):

\begin{itemize}
\item {\bf Volume of a geodesic ball:}  If we have well-defined lengths, volumes, and geodesics,
the volume of a small geodesic ball scales as $r^d$, where $d$ is the dimension.
\item {\bf Box-counting and Hausdorff dimensions:} To generalize the idea of measuring the scaling of 
volume, we  can cover a region by balls of diameter $\epsilon$ (box-counting) or diameter $\le\epsilon$ (Hausdorff), 
and determine the rate at which the required number of balls changes as $\epsilon$ shrinks.  The resulting
dimensions can be scale-dependent, and need not be integers.
\item {\bf Random walks:} Consider a random walk, or a diffusion process, in a region.  
The higher the dimension, the slower the walker will move away from the starting point, but the longer
it will take, on average,  to return to the starting point.  The rate of diffusion is determined
by a heat kernel, which, for a manifold of dimension $d$, has the asymptotic structure
\begin{align}
K(x,x';s) \sim (4\pi s)^{-d/2} e^{-\sigma(x,x')/2s} (1 + \mathcal{O}(s)),
\label{b1}
\end{align}
where Synge's ``world function'' $\sigma(x,x')$ is defined as  half the squared 
geodesic distance between $x$ and $x'$  \cite{Synge}, and
$s$ parametrizes the diffusion time.  The ``walk dimension'' is computed from the average 
distance as a function of $s$, while the widely-used ``spectral dimension'' is determined by the return 
probability, $K(x,x,s)$.   A causal version of spectral dimension, in which random walkers must move forward 
in time, can be defined in terms of the time for two random walkers starting at the same point to meet 
again \cite{Eichhorn}.
 \item {\bf Myrheim-Meyer dimension:}  In a Lorentzian spacetime, fix a causal diamond and measure
 the average size of subdiamonds contained within it.   Equivalently, for a discrete spacetime, count
 the number of points $C_1$ and the number of causal pairs $C_2$ in a causal diamond.  The ratio  
 \begin{align}
 \frac{\langle C_2\rangle\,}{\langle C_1\rangle^2} = \frac{\Gamma(d+1)\Gamma(\frac{d}{2})}{4\Gamma(\frac{3d}{2})}
 \label{b3}
 \end{align}
 determines a sort of box-counting dimension  that respects the causal structure.
 \item{\bf Behavior of geodesics:}  As an elaboration of Galileo's thought experiment, start with a
 spacetime that is ``really'' $d$-dimensional and follow $d-1$ orthogonal timelike geodesics starting
 from a given point.  In some spacetimes, the proper distance along such geodesics will be large in
 $d_{\hbox{\tiny IR}}-1$ dimensions and very small in the remainder.  The spacetime is then said to have 
 an effective ``infrared dimension'' $d_{\hbox{\tiny IR}}$ \cite{Hu}.
 \item {\bf Thermodynamic dimensions:} The density of states of a thermodynamic system---and, therefore,
 the partition function---depends on the phase space volume, which, in turn, depends on dimension.  As a
 result, many thermodynamic quantities have simple dependences on the space or spacetime dimension.
 The most straightforward measure comes from the equipartition theorem, which directly counts the
 translational degrees of freedom: For a monatomic gas,
 \begin{align}
 E \sim \frac{d-1}{2}NT.
 \label{b4}
 \end{align}
 However, the thermodynamic dimensions can also be obtained from the 
 temperature dependence of the free energy (this is the estimator used in \cite{Atick}), from the 
 Stefan-Boltzmann law, and from the equation of the state of a relativistic gas.
 \item {\bf Greens functions:} In one of the earliest investigations of dimension in physics, Ehrenfest
 pointed out that the Newtonian gravitational potential in $d$ spacetime dimensions scales as
 $r^{-(d-3)}$ \cite{Ehrenfest}.  This translates into a statement about Greens functions: The Hadamard 
 Greens function for a massless particle goes as 
 \begin{align}
G^{(1)}(x,x') \sim \left\{ \begin{array}{lc} \sigma(x,x')^{-(d-2)/2} \quad& d>2\\
                              \ln\sigma(x,x') & d=2 \end{array}\right. ,
\label{b2}
\end{align}
where $\sigma$ is again Synge's world function.   In many cases, though not always, this gives the same 
information as the spectral dimension, since when both are well-behaved the Greens function is a Laplace 
transform of the heat kernel (\ref{b1}). 
\item {\bf Scaling and anomalous dimensions:} Most physical quantities have natural ``scaling dimensions'',
which describe the way that they change under a constant rescaling of masses and lengths.  \mbox{These scaling}
dimensions, which can be determined classically by dimensional analysis, depend on the dimension of the
spacetime in which the fields are defined.  In quantum field theory, these canonical scaling dimensions can
receive quantum corrections---``anomalous dimensions''---that flow under the renormalization group.  As a
consequence, operators and observables in quantum field theory can act as if they live in a spacetime with a 
scale-dependent dimension.  \mbox{For example}, if the Greens function (\ref{b2}) is that of a quantum field theory,
the dimension $d$ is the full quantum-corrected and scale-dependent dimension, an effective quantum 
dimension of spacetime.  This is the dimensional estimator that most directly exhibits dimensional 
reduction in asymptotically safe gravity.
\end{itemize}

\section{What Is the Dimension of Spacetime?}

With this (incomplete) list of dimensions, we can now return to our original question: Does quantum gravity
exhibit dimensional reduction at very short distances?  The answer, of course, is that we don't know; we have
a plethora of definitions of dimension, which may or may not agree with each other, and we don't have a
complete, self-contained quantum theory to apply them to.  What we can do, though, is to collect evidence
from a number of different approaches to quantum gravity, to see if there is a pattern.  Again, readers
should consult \cite{Carlip3} for details.

\begin{itemize}
\item {\bf High temperature string theory:}  This is the setting in which, I believe, dimensional reduction was
first observed, using a thermodynamic dimension.  At high temperatures, a gas of strings undergoes a
phase transition in which the number of degrees of freedom drops dramatically, leading to, in the words
of Atick and Witten \cite{Atick}, a ``mysterious system'' that behaves at short distances ``as if this system
were a $(1+1)$-dimensional field theory.''  There is also some weaker evidence for dimensional reduction
coming
from the structure of high energy string \mbox{scattering \cite{Gross,Manes}}.

\item {\bf Causal dynamical triangulations:} The causal dynamical triangulations program is a discrete
gravitational path integral approach.  Much as a spherical geodesic dome is built from flat triangles, the curved 
manifolds of general relativity may be approximated by piecewise-flat simplicial manifolds, whose actions are 
then combined numerically to form a path integral.  Here, the most useful dimensional estimator is spectral
dimension, which can be defined without any smooth background manifold.  As first discovered by Ambj{\o}rn,
Jurkiewicz, and Loll in 2005 \cite{Loll}, this dimension falls from approximately four at large scales to
approximately two at small scales, with the reduction occurring at distances of about ten Planck lengths
\cite{Cooperman}.

\item {\bf Asymptotic safety:} As noted earlier, one of the most interesting predictions of asymptotically safe
gravity is dimensional reduction, again to $d=2$, near the ultraviolet fixed point.  The principal dimensional
estimator here is operator scaling dimension; it can be shown, on very general grounds, that near a nontrivial
UV fixed point, both gravitational and matter operators acquire large anomalous dimensions in such a way
that they scale as if they were in a two-dimensional spacetime \cite{Reuterb,Niedermeyer,Reuterc}.
Intuitively, this behavior occurs because the theory becomes scale-invariant at a renormalization group
fixed point, and it is only in two dimensions that the Einstein-Hilbert action is scale-invariant.  A similar
reduction to $d=2$ can be seen in the spectral dimension, \mbox{as determined} by the renormalized heat
kernel \cite{Reuterc}.

\item {\bf Short distance Wheeler-DeWitt equation:}  A similar sort of dimensional reduction can be found in 
the short-distance behavior of the Wheeler-DeWitt equation, although here it seems to have a different
origin \cite{Carlip2,Carlip4}.  In the short distance limit---or, equivalently, the strong coupling limit---the
coupling in the Wheeler-DeWitt equation between nearby spatially separated points becomes negligible 
\cite{Isham,Henneaux}.  This phenomenon, called ``asymptotic silence'' in cosmology \cite{Bruni,Andersson}, 
will be discussed in a bit more detail later.  For now, the important implication is that at very short distances, 
spacetime has an effective Kasner-like behavior.  Kasner space is four-dimensional, but in certain regimes
it acts two-dimensional: If one uses the behavior of geodesics as a dimensional estimator, one finds an
effective infrared dimension $d_{\hbox{\tiny IR}}=2$.

\item {\bf Loop quantum gravity and spin foams:}  The evidence for dimensional reduction in loop quantum
is mixed, but there are some indications that it occurs there as well.  As Modesto first noted \cite{Modesto},
the scaling of average area of a region changes at short distances, and an effective metric based on this
average leads to a spectral dimension that flows to $d=2$ near the Planck scale.  On the other hand, if
one defines the heat kernel directly on a spin network or spin foam, the flow of the spectral dimension depends
on the state \cite{Thurigen,Thurigen2}.  We probably need to know more about the space of physical states
before anything decisive can be said.

\item {\bf Noncommutative geometry/Snyder space:} The Snyder model was one of the first and simplest
models of a noncommutative geometry introduced in physics, and one of the first settings involving a 
fundamental minimum length \cite{Snyder}.  A variety of thermodynamic dimensions have been studied on
Snyder space \cite{Nozari}; all of them fall to $d=2$ at high temperatures.

\item {\bf Other noncommutative geometries:} 
Thermodynamic and spectral dimensions have also been studied in an assortment of other noncommutative 
geometries,
{{
with hints appearing as early as 2000 \cite{Lubo}.  The role of dimensional flow was emphasized
by Benedetti \cite{Benedetti}, who considered the spectral dimension in $\kappa$-Minkowski space;
the result was later extended to include a dimensional estimator based on the two-point function
\cite{Arzano}.  In more general noncommutative geometries, 
}}
a variety of different behaviors have 
been found \cite{Amelino}.  Noncommutative geometries often lead to modified dispersion relations,
which translate into modified heat kernels and, thus, modified spectral dimensions.  Unfortunately, this
setting is too general: {Any} energy dependence of the spectral dimension
can be reproduced from an appropriately-tuned modification of the dispersion relations \cite{Sotirou}.
For instance, a version of ``doubly special relativity''---a deformation of special relativity that combines
 frame independence with a new invariant energy scale---gives a spectral dimension that falls to 
 two at high energies, but the result depends on the choice of a free \mbox{parameter \cite{Gubitosi}}.  In the
 more mathematically formal noncommutative geometry of Connes \cite{Connes}, \mbox{it appears} that the
 spectral dimension is $d=0$ \cite{Alkofer}, but a variation leads to a spectral dimension that drops  
 from four at large distances to two at small distances \cite{Kurkov}.

\item {\bf Spacetimes with a minimum length:} It is plausible, although by no means certain, that the
quantization of spacetime will lead to a ``minimum length'', presumably on the order of the Planck
length.  A number of toy models have been built to explore this kind of behavior, and many of them
exhibit short distance dimensional reduction.  One can, for instance, modify the initial conditions for the
heat kernel (\ref{b1}) to describe diffusion of a wave packet with a minimum width \cite{Modesto2};
the resulting spectral dimension drops from $d$ at large distances to $d/2$ at small distances.  One
can introduce a slightly nonlocal ``quantum metric'' that incorporates a minimum length \cite{Chakraborty};
the box-counting dimension then drops to $d=2$ at small distances.   One can impose a ``generalized
uncertainty principle'' that changes the Heisenberg commutation relations to incorporate a minimum uncertainty
in length \cite{Husain,Hossenfelder}; this can also lead to dimensional reduction, although the details depend
on the exact assumptions.

\item {\bf Causal set theory:}  The starting point of causal set theory is a highly primitive 
representation of a discrete Lorentzian spacetime as a collection of points, characterized only by their 
causal \mbox{relations \cite{Sorkin}}.  The Myrheim-Meyer dimension, which was initially invented to describe 
causal sets, falls to approximately $d=2$ for small sets, in a fashion reminiscent of the 
short-distance behavior of the spectral dimension in causal dynamical triangulations \cite{Carlip1,Abajian}.  
The usual spectral dimension, on the other hand, {increases} at small distances \cite{Eichhorn}.
This may, however, merely reflect the fact that the d'Alembertian used to define the heat kernel is the 
``wrong'' one, one that fails to have the correct continuum limit.  The spectral dimension determined from
a corrected d'Alembertian shows the usual pattern of dimensional reduction to $d=2$ at short distances,
\mbox{as does} the associated Greens function \cite{Carlip1,Belenchia}.

\item {\bf Renormalizable modifications of general relativity:} Another possible approach to quantum gravity
is to modify the Lagrangian to make the theory renormalizable.  In one such approach, Ho{\v r}ava-Lifshitz
gravity \cite{Horava}, a generalized spectral dimension, flows to $d=2$ at high \mbox{energies \cite{Horava2}}, as 
do the thermodynamic dimensions \cite{Alencar}.  In curvature-squared models, the Greens function dimension
and a generalized spectral dimension exhibit a reduction to $d=2$ \cite{Gegenberg,Calcagni}.  In some
interesting nonlocal models, the spectral dimension also falls at high energies, although the exact behavior
depends on particular choices within the theory \cite{Modesto3}.
\end{itemize}
  
\section{What Is the Underlying Physics?}

As we have seen, quite a few indications point to to the existence of dimensional  reduction, most
likely to $d=2$, near the Planck scale.  One should treat this evidence with caution, though: While the same
phenomenon appears in many different approaches to quantum gravity, the {cause} of the dimensional
reduction seems to differ.  Unless one can find a common physical basis, it's not entirely clear how to
assess this evidence.  To be fair, the same problem occurs elsewhere in quantum gravity---we have many
different derivations of black hole entropy, for instance, but do not yet understand the underlying degrees of
freedom---but a physical picture would still be welcome.

We do not, of course, know the answer; if we did, this would be a very different review.  However, there are a
couple of recurring principles that are, at least, worth further investigation.

\subsection{Scale Invariance} 

Dimensional reduction in an asymptotically safe theory of gravity can be traced back to scale invariance
at the ultraviolet fixed point.  To see this, start with Newton's constant $G_N$, and define  the dimensionless 
coupling constant $g_N(\mu) = G_N\mu^{d-2}$, where  $\mu$ is the mass scale.  Under the RG flow, 
\begin{align}
\mu\frac{\partial g_N}{\partial\mu} 
   = [d-2+\eta_N(g_N,\dots)] g_N,
\label{d1}
\end{align}
where the anomalous dimension $\eta_N$ depends on both $g_N$ and any other coupling constants 
present in the theory.  Clearly, a free field (``Gaussian'') fixed point can occur at $g_N=0$.  At any
additional non-Gaussian fixed point $g_N^*$, however, the right-hand side of Equation (\ref{d1}) must vanish,
implying that $\eta_N(g_N^*,\dots) = 2-d$. 

But from the definition of the anomalous dimension, the momentum space propagator for a field 
with an anomalous dimension $\eta_N$ scales as $(p^2)^{-1 + \eta_N/2}$.  For $\eta_N = 2-d$, this 
becomes $p^{-d}$.  The position space propagator, obtained by a Fourier transformation, thus depends 
logarithmically on distance, which, from Equation (\ref{b2}), is characteristic of a two-dimensional conformal field.  A 
similar argument shows that any matter fields interacting with gravity at a non-Gaussian fixed point show 
the same two-dimensional behavior \cite{Niedermeyer}. The underlying physical explanation is simply that
Newton's constant is dimensionless and, therefore, potentially scale invariant, only in two spacetime 
dimensions. 

Scale invariance may also underlie dimensional reduction in causal dynamical triangulations.  In an
ordinary lattice field theory, the continuum limit is reached by tuning coupling constants
to a second-order phase transition.  At such a transition, the theory becomes scale invariant, and
correlations become long range, hiding the lattice structure \cite{Cardy}.  In causal dynamical triangulations, 
the situation is not yet completely clear, but there is growing evidence for a second-order phase transition
\cite{AGGJ}.  \mbox{There may,} indeed, be a direct connection to asymptotic safety: There are arguments
that causal dynamical triangulations may have a continuum limit only if the theory has an ultraviolet 
fixed \mbox{point \cite{Cooperman2}}.

Further hints of scale invariance come from the short distance limit of the Wheeler-DeWitt equation.  The 
full Wheeler-DeWitt equation (\ref{c1}) has two terms, which behave differently under scale transformations.
In the short distance/strong coupling limit, though, one term drops out, leaving an equation that is invariant
at least under constant rescalings.  

For other approaches to dimensional reduction, the situation
is less clear.  One might expect that theories with a fixed scale, such as those with a minimum length,
could not be scale invariant, but this would depend on how the length parameter flows under
the renormalization group.  For loop quantum gravity, a theory that has a minimum area, it
has been suggested that consistency conditions under coarse-graining may be similar to the asymptotic
safety requirement of a nontrivial ultraviolet fixed point \cite{Dittrich}.

\subsection{Asymptotic Silence}

A second, quite different recurring theme in dimensional reduction is that of ``asymptotic silence.''
A good starting point here is the Wheeler-DeWitt equation, 
\begin{align}
\left\{ 16\pi\ell_p^2G_{ijkl}\frac{\delta\ }{\delta q_{ij}} \frac{\delta\ }{\delta q_{kl}}
    - \frac{1}{16\pi\ell_p^2}\sqrt{q}\,{}^{(3)}\!R\right\}\Psi[q] = 0 ,
\label{c1}
\end{align}
where $q_{ij}$ is the spatial metric, $G_{ijkl}$ is the DeWitt supermetric, and $\ell_p = \sqrt{\hbar G/c^2}$ 
is the Planck length.  The wave function $\Psi[q]$ has structure at all scales, but we can focus in on small
distances by sending $\ell_p$ to infinity.  This is commonly viewed as the ``strong coupling'' limit, but it
is also the ``anti-Newtonian'' limit $c\rightarrow0$, in which the light cones collapse to lines.  Indeed, it is
evident  from Equation (\ref{c1}) that in this limit, \mbox{the only} term involving spatial derivatives, ${}^{(3)}\!R$, disappears, 
and nearby points decouple from \mbox{each other}.  

In cosmology, a similar phenomenon occurs near a spacelike singularity, where it is known as asymptotic
silence (nearby points can't ``hear'' each other) \cite{Bruni,Andersson}.  This is known to lead to BKL
behavior \cite{BKL}, in which spacetime looks locally like Kasner space, but with random axes and exponents 
that change under chaotic ``bounces.''  At short enough distances, Kasner space is certainly four dimensional,
but at somewhat larger (but, here, still Planckian) scales, it has an effective infrared dimension of two.
This reduction can also be seen in the spectral dimension.  The heat kernel for Kasner space at
(small) time $t$ is \cite{Futamase,Berkin}
\begin{align}
K(x,x;s) \sim \frac{1}{4\pi s^2}\left[ 1 + \frac{a}{t^2}\,s + \dots \right] .
\label{c2}
\end{align}

For a given $t$, one can always take the return time $s$ to be small enough that the first term dominates,
giving a ``microscopic'' spectral dimension of four.  However, for a given $s$---a fixed scale at which one is 
measuring the spectral dimension---there is always a time $t$ small enough that the second term dominates, 
giving a spectral dimension of two.

What makes this interesting is that asymptotic silence also appears in other approaches to dimensional
reduction.  There is evidence that quantum fluctuations of the vacuum generically induce 
collapsing light cones and asymptotic silence near the Planck scale \cite{Pitelli}.  Spacetime-like causal 
sets exhibit a behavior similar to asymptotic silence \cite{Eichhorn2}. Sorkin (cited in \cite{Eichhorn2}) 
has pointed out that asymptotic silence should lead to dramatically reduced scattering at high energies, 
a phenomenon seen in string theory \cite{Gross,Manes}. The reduction in the spectral dimension in 
formal noncommutative geometry may be a form of asymptotic silence \cite{Besnard,Bizi}, and quantum
corrections to the constraint algebra in loop quantum cosmology seem to also lead to high energy asymptotic 
silence \cite{Mielczarek}.   
  
Neither scale invariance nor asymptotic silence appear to be quite ``universal'': Each occurs in some version
of dimensional reduction, but is not apparent in others.  Nor do we understand whether the two phenomena are
related.  Still, they point out possible directions for finding a basic physical understanding of dimensional reduction.
  
\section{Can Dimensional Reduction Be Tested?}

Finally---since this is, after all, physics---we should ask whether there is any chance that dimensional reduction 
could be observed.  One's immediate impulse is to answer, ``Surely not'': The Planck scale seems very far
out of experimental reach.  However, remarkable progress has been made in measuring very tiny quantities,
and the question deserves some consideration.

{{Of course, such searches involve an implicit assumption that dimensional reduction affects the matter used
as a probe, and not just the gravitational sector.  For some dimensional estimators, this is clearly the case.  
The spectral dimension and the Greens function dimension are determined directly from the behavior of matter, 
as is the ``infrared dimension'' extracted from the characteristics of geodesics.  Anomalous dimensions can differ 
for different fields, but in asymptotic safety the reduction to $d=2$ occurs for matter as well as gravity 
\cite{Niedermeyer}.  Thermodynamic dimensions are less clear; they explore certain aspects of matter,
but one would have to check their relevance for particular experiments.  For other, more ``geometrical''
dimensional estimators, such as the Hausdorff dimension, \mbox{a more} concrete investigation of the physical implications
might be needed.

There are a number places to start to look for observable signals:}}

\begin{itemize}
\item {\bf Broken Lorentz invariance:} Searches for violations of Lorentz invariance have reached extraordinary
precision, ruling out effects that are suppressed by one or even two factors of the Planck mass \cite{Mattingly,Lor}.
Dimensional reduction to $d<4$ is surely incompatible with ordinary four-dimensional Lorentz invariance, so
one might hope that these tests could provide constraints.  Unfortunately, almost all existing tests have searched for 
so-called ``systematic'' breaking of Lorentz invariance---breaking with a single universal preferred direction.  Most
versions of dimensional reduction, on the other hand, involve ``nonsystematic'' breaking, in which the
preferred directions vary randomly in spacetime.  Although there have been searches for such effects
\cite{Lor,Basu}, they are generally much harder to see.
\item {\bf Laboratory tests:} Attempts to measure the Hausdorff dimension of spacetime at ``ordinary'' scales
date back to the mid-1980s, where measurements of the electron anomalous magnetic \mbox{moment \cite{Svozil}}
and the Lamb shift \cite{Shafer} were used to place fairly tight limits on deviations from $d=4$.  \mbox{Such 
measurements }do not probe the Planck scale, though, and even higher loop effects are fairly insensitive 
to Planck-scale dimensional reduction \cite{Shevchenko}.  The tightest current bounds are on a class of 
multifractional models, where particle physics measurements restrict the characteristic length scale for
dimensional reduction to $\ell_*<10^{-17}\,\mathrm{m}$ or, for certain parameters,  $\ell_*<10^{-27}\,\mathrm{m}$
\cite{Calcagni2}.  Such measurements are closing in on the Planck scale, but the limits are rather 
model-dependent.

\item {\bf Cosmology:} Cosmology may provide the best opportunities for testing dimensional reduction.  The
expansion of the Universe, especially in an inflationary period, stretches out Planck-scale signals to observable
sizes.  Here, the problem is to disentangle the evidence of dimensional reduction from other phenomena.  For
example, as Amelino-Camelia et al. have emphasized, a reduction to two dimensions in the early Universe
would naturally lead to a nearly scale-invariant spectrum of fluctuations of the cosmic microwave
background, a signal usually interpreted as evidence for inflation \cite{Amelino2,Amelino3}.   Work has
begun on exploring the cosmological implications of asymptotic \mbox{safety \cite{Bonanno}}, along with several other
models of dimensional reduction \cite{Brighenti,Calcagni3,Mielczarek2}; these may ultimately lead to observable 
consequences.
\end{itemize}

\section{Conclusions}

Does quantum gravity lead to dimensional reduction near the Planck scale?  We don't know; for now, we have neither 
the theoretical apparatus nor the experimental observations to give a firm answer to this question.  However, there 
are intriguing indications, coming from enough different approaches to quantization to make this a serious
question.  

If dimensional reduction {does} occur, it will have deep implications for physics, from the small scale 
structure of spacetime to cosmology.  It may reveal new symmetries, such as scale invariance, 
or new phenomena, such as short-distance asymptotic silence; it might even tell us that our four observed
dimensions are somehow ``emergent'' from a two-dimensional spacetime.  It will be interesting to see how
the subject develops.

\begin{flushleft}
\large\bf Acknowledgments
\end{flushleft}

This work was supported in part by Department of Energy grant DE-FG02-91ER40674.

\end{document}